# Photon-number-resolution with sub-30-ps timing using multi-element superconducting nanowire single photon detectors

Eric A. Dauler[a,b*], Andrew J. Kerman[b], Bryan S. Robinson[b], Joel K. W. Yang[a], Boris Voronov[c], Gregory Gol'tsman[c], Scott A. Hamilton[b], and Karl K. Berggren[a]

[a]*Massachusetts Institute of Technology, Cambridge, MA 02139, U.S.A.;* [b]*Lincoln Laboratory, Massachusetts Institute of Technology, Lexington, MA 02420, U.S.A.;* [c]*Moscow State Pedagogical University, Moscow 119345, Russia.*

A photon-number-resolving detector based on a four-element superconducting nanowire single photon detector is demonstrated to have sub-30-ps resolution in measuring the arrival time of individual photons. This detector can be used to characterize the photon statistics of non-pulsed light sources and to mitigate dead-time effects in high-speed photon counting applications. Furthermore, a 25% system detection efficiency at 1550 nm was demonstrated, making the detector useful for both low-flux source characterization and high-speed photon-counting and quantum communication applications. The design, fabrication and testing of this detector are described, and a comparison between the measured and theoretical performance is presented.

*Corresponding author. Email: edauler@mit.edu

# Photon-number-resolution with sub-30-ps timing using multi-element superconducting nanowire single photon detectors

A photon-number-resolving detector based on a four-element superconducting nanowire single photon detector is demonstrated to have sub-30-ps resolution in measuring the arrival time of individual photons. This detector can be used to characterize the photon statistics of non-pulsed light sources and to mitigate dead-time effects in high-speed photon counting applications. Furthermore, a 25% system detection efficiency at 1550 nm was demonstrated, making the detector useful for both low-flux source characterization and high-speed photon-counting and quantum communication applications. The design, fabrication and testing of this detector are described, and a comparison between the measured and theoretical performance is presented.

# 1. Introduction

The term "photon-number-resolving" is typically used to describe a photon-counting detector that can resolve the number of simultaneously-incident photons, even if the detector cannot differentiate between photons incident simultaneously and those closely-spaced in time. Photon-number-resolution without precise timing is useful in many research efforts, including linear optics quantum computing [1], conditional state preparation [2], and source characterization for enhanced quantum-key-distribution security [3-4]. However, photon-number-resolving detectors with precise timing could benefit other applications that employ photon-counting detectors, such as high-sensitivity optical communication [5], laser radar [6], and fluorescence measurement techniques [7], particularly if the ability to resolve and time multiple photons allowed dead-time effects to be mitigated. Unfortunately, most proposed and demonstrated photon-number-resolving detectors only resolve photon number at the expense of high timing jitter and low counting rate. A photon-number-resolving detector with speed and timing resolution matching or outperforming the best available for single-photon detectors would be enabling for many systems requiring high-speed performance.

The first of two approaches to achieving photon-number-resolution is to use a linear-mode detector with sufficiently low noise that the number of simultaneously-detected photons can be resolved by measuring the amplitude of the resulting output signal. Several detector technologies capable of resolving photon number in this way have been demonstrated including visible light photon counters [8], superconducting transition edge sensors [9], and superconducting tunnel junction detectors [10]. The fidelity with which a photon-number measurement can be made using these devices is limited by both non-unity detection efficiency and the signal-to-noise ratio of the output signal. This fidelity becomes progressively worse in distinguishing *n* from *n+1* photons as *n* becomes large. Although the detection efficiency for many of these technologies is high and they are excellent for measuring the photon-number statistics of low-rate, pulsed sources, measuring non-pulsed sources is more challenging. If the photons do not arrive in pulses well-separated in time, resolving detection events that occur while the detector is recovering from a previous event becomes a complex problem, and the measurement result can depend strongly on not only the detector, but also the readout electronics.

The second approach for achieving photon-number-resolution is to split the light between many spatial or temporal modes, so that each mode contains much less than 1 photon on average, and to measure each of these modes separately using single-photon-sensitive detectors whose outputs can be digitally combined. This approach has been proposed and demonstrated both for spatial multiplexing [11-12] and for temporal multiplexing [13-14]. Using the multiplexing approach, the fidelity of the photon-number-resolution measurement is again limited in part by the detection efficiency. However, in contrast to the linear-mode detector approach, the second contribution to the measurement fidelity is imposed by the number of modes in which the light is detected, not by the signal-to-noise ratio of the detectors. Although spatial multiplexing requires more detector elements than temporal multiplexing, spatial multiplexing *increases* the maximum counting rate, is compatible with non-pulsed light sources and preserves the timing resolution achievable with the individual single-photon detectors. This paper will describe how spatial multiplexing may be implemented using superconducting nanowire single photon detectors to achieve photon number resolution, high counting rates and excellent timing resolution, without incurring the insertion losses or high dark count rates that impacted previous spatially-multiplexed implementations.

## 2. Detector concept

Single-photon-sensitive detectors, which are capable of detecting a single photon, but cannot resolve whether one or several photons are simultaneously incident, can only reliably measure photon number for states with zero or one photon. In order to use these detectors to measure photon number for states with more than one photon, we may use an N-port device to split one mode into many, each with fewer photons on average than the original state [11]. We can then use single-photon-sensitive detectors to measure the photon number in each output mode of the N-port (Figure 1(a)) and digitally add the number of detector counts. In the case of spatial multiplexing, each output mode is coupled into a spatially-distinct, independent detector element. Several implementations of an N-port could be considered for spatial mutliplexing, such as multiple optical beam-splitters in a tree arrangement or an array of detector elements across which the optical beam is spread. In this paper, a unique approach to tightly packing the detector elements will be presented in which the packing occurs on a sub-diffraction-limited length scale. In evaluating this approach for implementing an N-port, the important considerations are: (1) vacuum modes at all inputs except for the one of interest, (2) sufficiently low loss, (3) enough detector elements that the probability of any detector having more than one photon is sufficiently low and (4) detectors with high-efficiency and low-noise.

Given all of the requirements above, it is useful to consider what applications may benefit from the spatial multiplexing approach. One often cited advantage of the multiplexing approach in general is that it can be used to transform any single-photon-sensitive detector technology into a photon-number-resolving detector, including technologies that operate near room temperature. Applications requiring compact photon-number-resolving detectors may benefit from this approach, which has been demonstrated for both spatial [12] and temporal [13-14] multiplexing. However, one advantage unique to spatial multiplexing is the ability to create a photon-number-resolving detector with higher performance, particularly in terms of the maximum counting rate and timing resolution. These performance improvements may be realized both because a detector technology may be selected specifically for its speed and because many independent detector elements will count in parallel. Thus, for applications requiring high-speed and precise timing resolution, the spatial multiplexing approach is preferable.

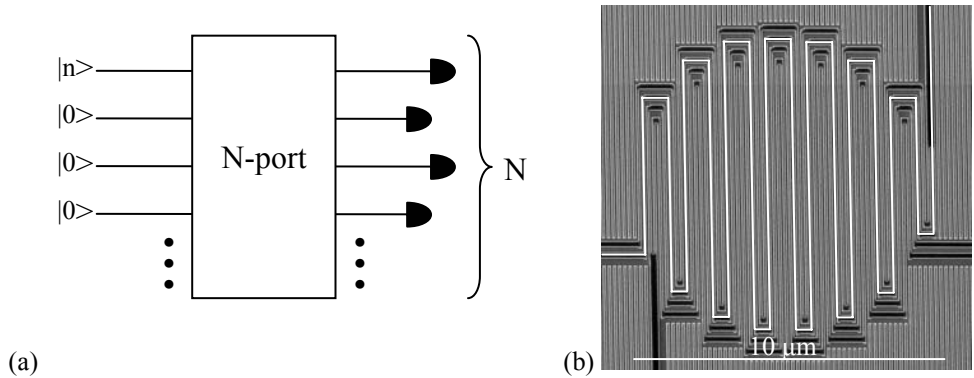

Figure 1. (a) N-port concept for multiplexing single-photon-sensitive detectors to construct a photon-number-resolving detector and (b) scanning electron microscope micrograph of a 4-element SNSPD with highlighting to indicate one of the four independent elements

Superconducting nanowire single-photon detectors (SNSPDs) are ideal for use in a spatially-multiplexed, photon-number-resolving detector designed for high-speed performance because of their low timing jitter and high maximum counting rate. These detectors are composed of ~5-nm-thick, 50-to-100-nm-wide superconducting wire that is biased at a current density just below the threshold at which the material switches from the superconducting to the resistive state [15]. In this mode of operation, the absorption of a single photon is sufficient to cause a resistive bridge to form across the nanowire, resulting in a measurable voltage pulse. This resistive state is unstable in an SNSPD and the nanowire quickly returns to the superconducting state following a detection event [16]. The current recovery, or dead-time, is limited by the kinetic inductance of the superconducting wire, which is proportional to the wire length, and is typically several nanoseconds [17]. Finally, the timing jitter can be < 30 ps [18].

Multiple superconducting nanowire single photon detector elements can be arranged, without requiring any optical elements, in order to function as an N-port that satisfies all of the requirements previously mentioned and to improve the detector performance. Figure 1(b) illustrates one arrangement of four SNSPD elements, which are interleaved so that the NbN nanowires forming the active area are nearly identical to those of a single SNSPD covering the same active area [18]. In this way, there are no losses associated with splitting the light between the elements, which degrades the performance of other approaches for multiplexed detectors. Furthermore, there are no new propagating optical modes that must be assured of having vacuum inputs; the input to the multi-element SNSPD is identical to that of a single-element SNSPD because they can have identical active areas. The fact that the total detector area is fixed also limits the effect of other area-dependent sources of degradation, including the degradation of detection efficiency due to increase probability of a defect [19], reset time [17] and possibly dark counts with increasing device area. In fact, the multi-element approach actually allows smaller individual elements to be used, improving the overall detection efficiency [19] and reset time [18].

The remaining challenge to making the multi-element SNSPD practical as a photon-number-resolving detector is reading out a sufficient number of independent elements. Ideally, the detector elements would be connected to an integrated circuit capable of providing readout for a large number of elements, as has been demonstrated for Geiger-mode avalanche photodiodes [12]. While work on such a readout-integrated circuit progresses, reading out a small number of elements using discrete electronics is feasible. The next section will examine the question of how many elements are needed for a useful photon-number-resolving detector.

## 3. Theory

In this section, we will examine the photon-number-resolution measurement fidelity for two reasons. First, we will determine under which conditions this fidelity is limited by the finite number of elements, as opposed to the detection efficiency, in order to determine whether it is more important to focus on developing a detector with the highest number of elements or the highest detection efficiency possible. Second, we will calculate the count statistics we would expect to measure using a spatially-multiplexed detector in order to provide a way of evaulating the performance of the detector later in this paper.

### *3.1 Photon-number-resolving fidelity limitations*

As mentioned above, the fidelity of a photon-number measurement using the spatial multiplexing approach is limited by both the detection efficiency and the number of single-photon-sensitive elements composing the detector. It is important to isolate each of these effects in order to determine the circumstances in which the number of elements is the dominant limitation. We will show in the following section that by using only 4-elements, the achievable SNSPD detection efficiency is the dominant limitation to the fidelity of measurements of 1, 2, 3 or even 4 photons. The theory will also provide us with a method for evaluating the performance of the 4-element SNSPD and for predicting what improvements in the measurement fidelity may be expected by increasing the number of elements or improving the detection efficiency. In order to simplify the discussion, the detection efficiency will be defined to include all losses incurred in coupling the light onto the multi-element detector. The probability of measuring the correct number of photons in a pulse with $n$ photons, $P(n|n)$, has previously been found analytically for the case of equal splitting between $N$ detector elements, each with the same detection efficiency $\eta$ [13]:

$$P(n|n) = \frac{N!}{N^n(N-n)!}\eta^n \qquad (1)$$

This expression was used to perform the calculations in the remainder of this section. The solution to Equation 1 is plotted as a function of detection efficiency and the number of detector elements, where Figure 2(a) shows the probability $P(2|2)$, Figure 2(b) shows the probability $P(3|3)$, and Figure 2(c) shows the probability $P(4|4)$.

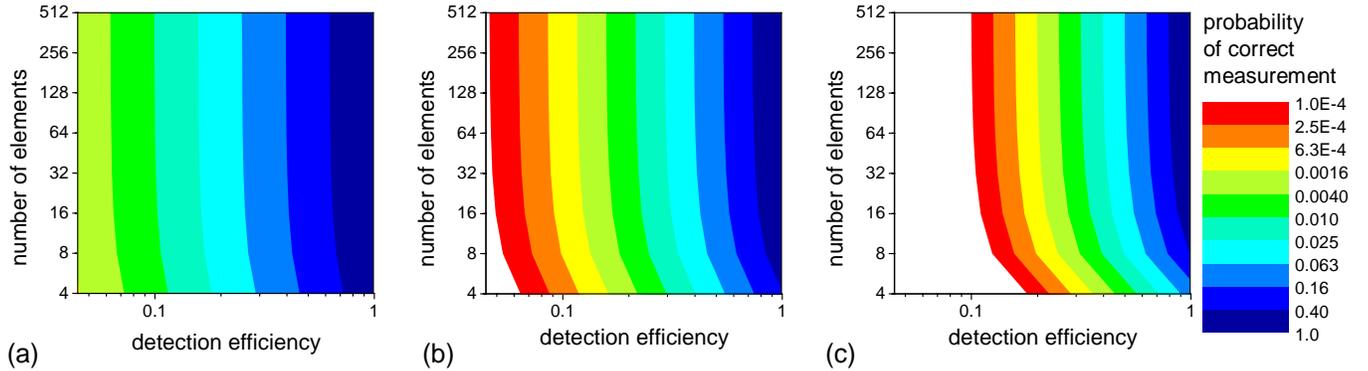

Figure 2. Calculated probabilities (a) $P(2|2)$, (b) $P(3|3)$, and (c) $P(4|4)$ for a spatial multiplexed detector plotted as a function of the number of elements and the detection efficiency, which is the same for all of the elements between which the light is equally split. Note that the contours of equal measurement fidelity are logarithmically spaced.

Both Equation 1 and Figure 2 illustrate that the measurement fidelity is scaled by a factor associated with the detection efficiency and that reduced detection efficiency rapidly degrades the multi-photon measurement fidelity. The measurement fidelity is also reduced by a fixed factor associated with the number of elements. In the case of measuring zero or one photon, the scale factor is unity, while for measuring more than $N$ photons ($n > N$), where $N$ is the number of elements, this scale factor is zero. Consequently, for a given number of elements, the fidelity for measuring $N$ of fewer photons is limited primarily by the detection efficiency when this efficiency is low and by the number of elements when the detection efficiency is high. The fact that the terms containing the detection efficiency and the number of elements in Equation 1 are separable allows us to easily compare the penalty from each. The penalty associated with the detection efficiency is simply $\eta^n$ and we will define $\alpha_{DE} = 10\ log_{10}(\eta^n)$ as this penalty in decibels. Similarly, the penalty for using a finite number of elements is given by the other factor in Equation 1 and we will define $\alpha_N$ as this penalty in decibels. The penalty due to the number of elements for measuring one photon is always 0 dB, because the photon will be detected with the same probability regardless of the number of elements to which it is directed. For measuring two photons, the penalty is 1.25 dB (25%) for using $N=4$ elements and only 0.28 dB (6.3%) for using $N=16$ elements. Similarly, for measuring three photons, the penalty is 4.26 dB (62.5%) for using $N=4$ elements and 0.86 dB (18%) for using $N=16$ elements.

Although these penalties are significant, they must be compared to the penalties associated with the detection efficiency, $\alpha_{DE}$. This comparison is made in Figure 3, where the ratio of the penalty from the number of elements, $\alpha_N$, to the total penalty, $\alpha_{DE} + \alpha_N$, is plotted as a function of the detection efficiency. The detection efficiency at which the penalty from the number of elements dominates ($\alpha_N < \alpha_{DE}$) can be easily seen in Figure 3(b), where the horizontal axis intersects the vertical axis at $\alpha_{DE} = \alpha_N$. For the commonly encountered case of detecting two photons, the detection efficiency of the elements must be high (>86.6% for the case of $N = 4$ and >96.8% for the case of $N = 16$) in order for the penalty from the number of elements to dominate. In fact, even for the case of detecting three or four photons with a 4-element detector, the detection efficiency penalty dominates when the efficiency is below 72.1% or 55.3% respectively. Given that the detection efficiency and coupling losses of the superconducting nanowire single-photon detector elements will be the primary limitation to its performance for measuring four or fewer photons, there is little fidelity to be gained by using more than four elements in applications that do not require counting large numbers of photons. Therefore, a four-element superconducting nanowire single-photon detector will be used to demonstrate the photon-number-resolving capability of this approach.

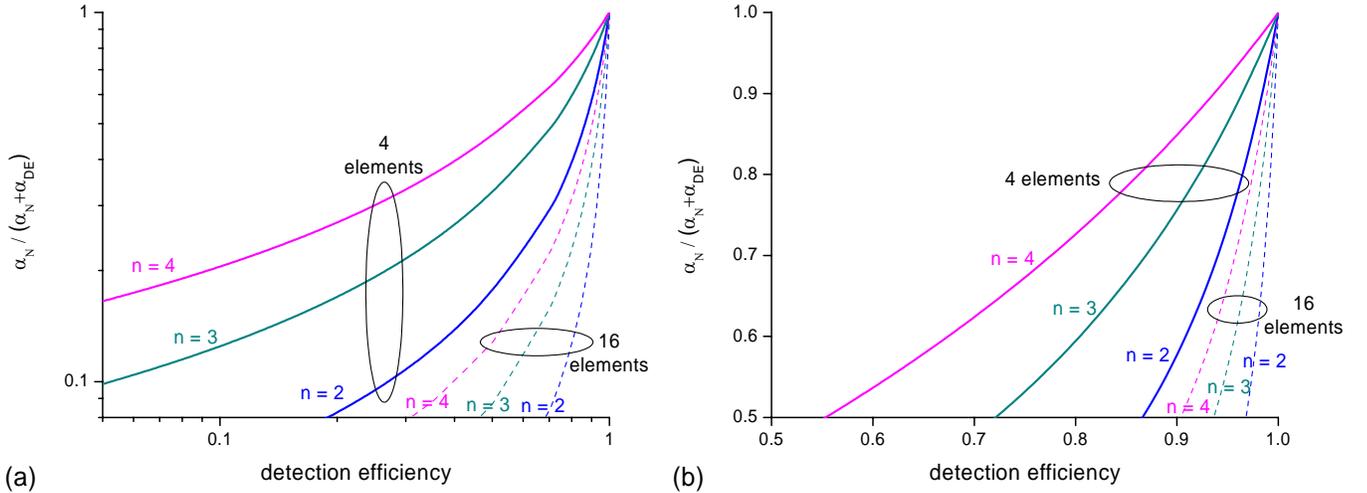

Figure 3. Fraction of the measurement fidelity penalty due to the finite number of elements plotted on (a) logarithmic and (b) linear scales. The detection efficiency must be greater than the value at which the curves in (b) intersect the horizontal axis ($\alpha_{DE} = \alpha_N$,) in order for the penalty from the number of elements to dominate.

## 3.2 Photon-number statistics for coherent light measured with an N-element detector

Although we have now examined the fidelity of the spatially-multiplexed approach in resolving photon number states, it is more straight-forward to test the detector with a coherent source. We may again use the probabilities previously calculated in reference 13, along with the photon statistics for coherent light to find the probability, $P^N_\eta(m|\lambda)$, of measuring $m$ counts from an optical pulse with, on average, $\lambda$ photons using an N-element detector with detection efficiency $\eta$:

$$P^N_\eta\left(m|\lambda\right) = \sum_{n=m}^{\infty} \frac{N!}{m!(N-m)!} \frac{(\eta\lambda)^n e^{-\eta\lambda}}{n!} \sum_{j=0}^{m}(-1)^j \frac{m!}{j!(m-j)!}\left(1 - \eta + \frac{(m-j)\eta}{N}\right)^n \quad (2)$$

This result will provide a way to compare measurements made with a coherent source to what can be theoretically expected.

## 4. Experimental demonstration

Validating the proposed approach for simultaneously achieving photon-number-resolution and precise-timing-resolution consisted of several steps that we will describe in this section. First, we will discuss the design and fabrication of the 4-element SNSPD. Next we will describe how the individual elements were tested and how an appropriate device was packaged so that all four element could be operated simultaneously with a high system detection efficiency. Then, in order to test the abilities of the detector, two measurements were performed. The first measurement was to characterize the photon-number-resolution using a short-pulsed optical source in order to verify that the measured photon-number statistics closely match Equation 2. Finally, the precise timing resolution was demonstrated at the same time as the photon number resolution by measuring the photon statistics from two short optical pulses spaced by 100 ps in time.

### 4.1 Design and fabrication

The design of the spatially-multiplexed superconducting nanowire single photon detector must take into account several desirable properties discussed above, including: (1) a sufficient number of elements, (2) equal splitting of the light between the elements and (3) elements with as high detection efficiency as possible. First, we found in section 3.1 that four elements suffice for measurements of up to four photons, but do not introduce too much complexity in the packaging and readout. Second, interleaving the four elements, as mentioned in section 2 and shown in Figure 1(b), is an ideal design for ensuring the light is split equally between the elements and for making the splitting ratio insensitive to the optical alignment of the detector. Finally, maximizing the detection efficiency of the elements requires several tradeoffs, which will be discussed in detail in a future publication. Briefly, this optimization must consider all three contributions to the detection efficiency: the coupling efficiency, the absorption and the internal detection efficiency. Tradeoffs exist between the device area, wire width, fill factor and film thickness, which for the detector used in this work were an approximately circular active area with ~9.4-μm diameter, 90-nm wire width, 56% fill factor and ~5-nm NbN film thickness. Furthermore, an optical cavity [20] was fabricated on top of the devices to enhance the absorption at 1550 nm wavelength without sacrificing the other contributions to the detection efficiency.

The detector structures were fabricated at the Massachusetts Institute of Technology using a previously reported process [21], with several modifications, on ultrathin (~5-nm) NbN films deposited at Moscow State Pedagogical University [22]. The fabrication process consists of 2 stages: (1) photolithography, Ti/Au evaporation and liftoff steps to define electrical contact pads and (2) electron-beam lithography and reactive-ion etching steps (using the electron-beam resist as the etch mask) to pattern the NbN into the detector structure. In this work, the hydrogen silsesquioxane used as an electron-beam resist for defining the detector structure was XR-1541 (6% solids) and was spun at 6.5 krpm to a thickness of 80 nm. The detector pattern was defined as a series of rectangles with 100-nm width, 10-nm wider than the etched, 90-nm, NbN-nanowire width. The actual width of the patterned wire depends not only on the exposure width, but also on the development process and the exposure dose. The electron-beam dose is not selected to match the wire width to the exposed feature width, but is instead selected to minimize the roughness of the feature edges. The detector patterns were exposed at an electron-beam dose ~10% lower than the dose at which footing, thin layers of residual resist in unexposed regions at the edges of exposed resist, first became visible. Proximity-effect correction was achieved by exposing dummy rectangles with the same width and pitch as the detector structure, but disconnected from the detector, around its perimeter. The dummy structure extended ~ 5 μm from the edges of the detector, and although it introduced more background dose than might be achieved with an optimal design, it required no computation to create and provided good linewidth uniformity across the detector active area. The electron-beam writing was performed using a 30-μm aperture and a 30-kV accelerating voltage, resulting in a 230-pA beam current. The resist was developed by immersing the chip in 25% tetramethyl-ammonium-hydroxide for 4 minutes. The chip was agitated by hand after 2 minutes to remove bubbles from the surface. Patterning errors, generated by the electron-beam writer, resulted in an offset of 20 – 50 nm between patterns written from horizontally and those from vertically-oriented rectangles in the pattern file, although this did not measurably affect the pattern in the active region of the device and was compensated when necessary to avoid gaps along the NbN wire.

The cavity structures were also added using a previously reported process [20], with modification of the dielectric. The thickness of the dielectric spacer, which is formed by e-beam exposing a second, thicker layer of hydrogen silsesquioxane in a large region covering the detector, is complicated by several factors. These factors include the residual hydrogen silsesquioxane used to pattern the detectors, the local topography of the surrounding electrical contact pads, the electron dose used to expose the cavity dielectric and the development process used following the exposure. Instead of carefully characterizing these factors, the required thickness of the hydrogen silsesquioxane layer was overestimated (spun to 260 nm on a bare substrate). The cavity spacers were exposed with the electron-beam, developed and were measured using a Dektak 3 surface profilometer to determine their thickness over actual detectors (measured to be 230 nm). The chip was then etched using the same $CF_4$-based process used for defining the detectors to achieve the desired, 210 nm, spacer thickness.

*4.2 Packaging and testing*

Initial testing of the detectors was performed in a cryogenic probing station, as described previously [19, 21]. This testing was conducted primarily to select a 4-element detector in which the elements had high and roughly equal detection efficiencies. It was also used to calibrate the device detection efficiency, subtracting for coupling losses, since it is straightforward to measure the size of the optical beam in this setup. Once an appropriate 4-element detector was selected, the devices were protected with a ~1.5-μm-thick layer of Microposit S1813 photoresist and a ~4-mm square section of the chip centered on this detector was diced from the full chip.

A second setup, based on a closed-cycle cryocooler, was used to focus most of the light from a single-mode optical fiber onto the detector and to operate all four detector elements simultaneously. The chip was mounted using silver paint on an Au-plated copper holder with a ~2-mm-diameter hole behind the selected detector. The light was focused through this hole and the sapphire substrate onto the detector using a lens assembly fixed to the fiber, all attached to a 3-axis nanopositioner (Attocube ANPxyz50). The high-speed output signal of each element was read out through a separate coaxial cable and amplified outside of the cryocooler. The DC current bias was supplied by four separate battery-powered voltage sources, connected to the devices through cooled, 100-kΩ resistors located on the same mount as the detector and wirebonded to the signal line connected to each element.

The detector performance can be compared between the two setups to determine the coupling efficiency and to provide a basis for evaluating its photon-number-resolving capability. The device detection efficiency, determined by dividing the sum of the counting rate from all four elements by the incident number of photons within the 9.4-μm-diameter active-area of the device, was measured in the probing station to be 40% at 95% of $I_c$ at 2.7 K. The system detection efficiency, determined by dividing the sum of the counting rate from all four elements by the number of photons at the fiber input to the setup, was measured under the same conditions in the closed-cycle cryocooler setup to be 25%. Thus, the coupling efficiency, including losses in the fiber, lens assembly and light not focused inside the detector active area, is 63%. This relatively low coupling efficiency is likely due to the working distance being too short to allow the spot to be fully focused on the detector, as it was not possible to translate the positioners such that the count rate decreased beyond the point at which the maximum system detection efficiency was achieved.

*4.3 Photon-number-resolution*

We may now compare the photon-number-resolving ability of the 4-element SNSPD to the theory discussed in section 3.2. The output from each of the four detector elements was recorded as the attenuation of the 1550-nm-wavelength, mode-locked source was varied to produce optical pulses with between ~0.01 and 20 photons per pulse. The electrical output from the amplifiers connected to each detector element was connected to a single channel on a high-speed digital oscilloscope. Additionally, the oscilloscope was triggered by an input locked to the frequency of the mode-locked laser. The oscilloscope was then set up to record a sequence of 10,000, 5-ns traces for each of the four channels. In this way, the simultaneous output from all four detector elements could be recorded and only the electrical output of the detectors within a 5-ns period centered on the timing of the optical pulses had to be stored and analyzed. At each attenuation setting, between 5 and 10 sequences were recorded, for a total of between 50,000 and 100,000 optical-pulse periods. These files were then post-processed in Matlab to extract the times at which each detector-element output crossed a fixed threshold. Using these times, the number of detector elements that fired within ±50 ps of an optical pulse (set independently for each channel to account for differences in the propagation time between the detector element and the oscilloscope) was calculated for each optical pulse period.

The probability of measuring zero, one, two, three or four counts was calculated from the data for each attenuation. These probabilities are plotted as the markers in Figure 4(a), with the horizontal axis scaled by the measured, low-flux detection efficiency (i.e. the incident flux was actually 1/DE = 4 times higher). Additionally, the theory discussed in section 3.2 was used to predict the photon statistics, using $N = 4$ elements and assuming the light divided equally between the elements, with each element having the average measured detection efficiency. The small variations between the measured individual-element detection efficiencies (section 4.2) and the interleaved geometry justify this assumption. These theory curves are plotted as lines in Figure 4(a), and the excellent agreement with the measured count statistics confirm that the detector provides the expected photon-number-resolution.

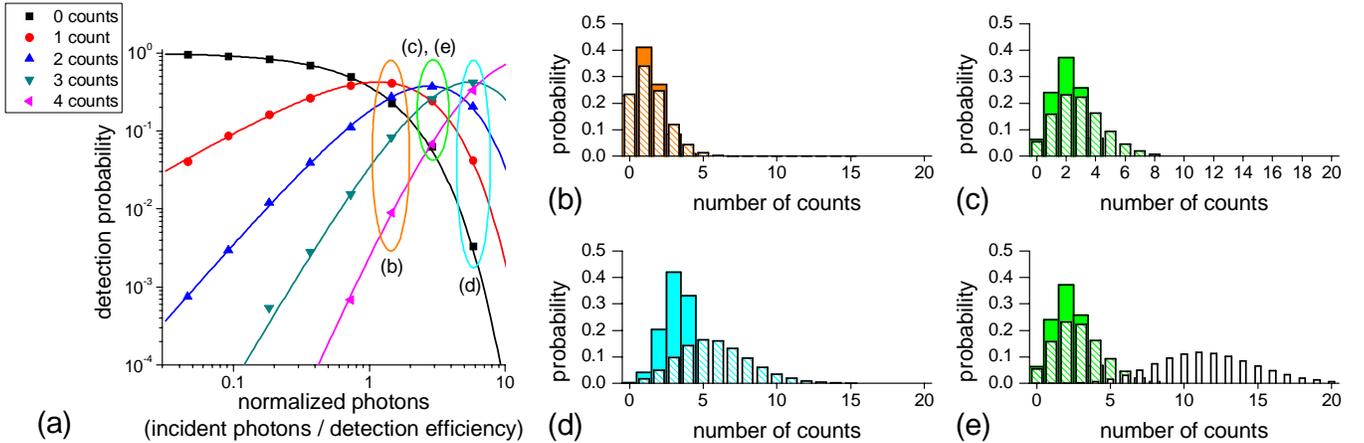

Figure 4. (a) Measured (data points) and calculated (lines) detection probabilities for each possible number of counts within a 100 ps timing window as a function of the normalized photon flux. The measured data points are plotted as a function of photon flux normalized by the measured detection efficiency and the calculations are for a four-element detector with unity detection efficiency. The measured detection probabilities (solid filled bars) are also plotted as a function of the number of counts in (b), (c), (d) and (e) for three values of photon flux. Additionally, calculations are shown (hashed bars) assuming $\alpha_N = 1$ (no penalty from a finite number of elements) and $\alpha_{DE} = \eta^n$, where $\eta$ is the measured detection efficiency. The penalty from using 4 elements can be compared to the penalty from the non-unity detection efficiency in (e), where the calculations assuming $\alpha_N = \alpha_{DE} = 1$ are also shown

(white bars). It is clear that for counting 4 or fewer photons, the fidelity of the photon-number measurement is primarily limited by the detection efficiency and not by the number of elements.

Finally, it is useful to compare the measured count statistics to those that would be expected both for a photon-number-resolving detector without a limited number of elements and for an ideal detector (i.e. also having unity detection efficiency). The comparison to a detector without a limited number of elements is made for the three highest measured photon fluxes in Figures 4(b)-(e) and the comparison to an ideal detector is made in Figure 4(e). It is clear that the effect from having only four elements is small compared to the effect of the non-unity detection efficiency.

*4.4 Mitigated dead-time effects and precise timing resolution*

Although the photon-number-resolution of the 4-element SNSPD has now been demonstrated, its primary advantage over other photon-number-resolving detectors is its ability to precisely time each photon detection event, even for non-pulsed optical signals. This ability makes the multi-element SNSPD useful for measuring the photon-statistics as a function of time for high-speed sources or for mitigating dead-time effects in a variety of applications [6-8].

To investigate these abilities using our four-element detector, we repeated the experiment performed in the previous section, but with two optical pulses spaced by 100 ps in time. The readout and data analysis was carried out in the same way, but instead of only counting the number of detectors that fired in a single 100 ps period, we counted the number of detectors that fired in several, variable-length, time periods. Figure 5(a) shows the probability of measuring different numbers of counts versus time for 100 ps time bins while figure 5(b) shows the same for 12.5 ps time bins. Virtually all of the detection events resulting from a single optical pulse can be collected in a 100-ps time bin. Additionally, when the time bins are shorter than the ~30-ps-FWHM timing jitter of the detector, the photon statistics can be measured on the time scale of this jitter. In fact, careful measurements of the detector jitter using a short-pulsed optical source such as the mode-locked laser used in this work could allow the photon statistics to be measured on a sub-30-ps time scale by deconvolving the detector jitter from the data. However, using short time bins result in the detection events being spread across several bins so that the number of detection event in any given bin is lower (i.e. it acts like an additional loss), so the improved timing resolution comes at the expense of requiring more data. Recording the time stamp of each detection event allows the data to be post-processed in the most appropriate way to obtain the optimal trade-off between precise timing resolution, low uncertainty of the photon-number statistics and short data collection times.

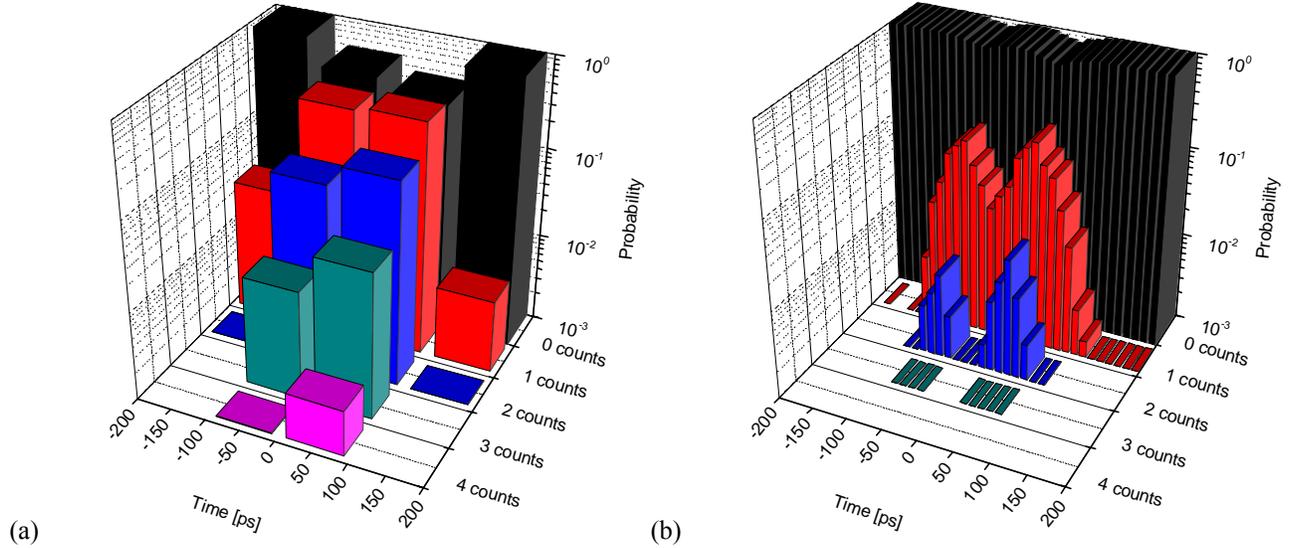

(a)  (b)

Figure 5. Measured photon number statistics (detection probability for measuring each number of counts within the timing window) as a function of time delay for two optical pulses separated by 100 ps in time. The same data is plotted after software analysis is used to determine the statistics for (a) 100 ps time windows and (b) 12.5 ps time windows. The two optical pulses can be clearly resolved in (b) and the photon number statistics for each pulse can be more quickly measured (out to four simultaneous counts) using the wider time windows shown in (a).

## 5. Conclusion

An approach for achieving photon-number-resolution with precise timing information on each detection event has been demonstrated using a 4-element superconducting nanowire single-photon detector. The fidelity of a photon-number-resolution measurement made using such a detector has been analyzed, providing support for using only four elements and verifying that the demonstrated detector operates as expected. A 25% system detection efficiency was achieved and the unique timing capabilities of this photon-number-resolving detector were demonstrated, including the ability to measure photon-number statistics with unprecedented timing resolution. Therefore, this detector is ideally suited for many applications requiring a high-speed detector that may also benefit from photon-number resolution or reduced dead-time effects.

## 6. Acknowledgements

The authors would like to thank Prof. H. I. Smith for the use of his facilities and equipment, Mr. J. Daley and Mr. M. Mondol for technical assistance, and Ms. L. Hill for wirebonding the detectors. This work made use of MIT's shared scanning-electron-beam-lithography facility in the Research Laboratory of Electronics (SEBL at RLE).

This work was sponsored by the United States Air Force under Air Force Contract #FA8721-05-C-0002. Opinions, interpretations, recommendations and conclusions are those of the authors and are not necessarily endorsed by the United States Government.